\documentclass[prb,twocolumn]{revtex4}
\usepackage{amsmath}
\usepackage{amssymb}

\usepackage[colorlinks=true,citecolor=blue,urlcolor=blue]{hyperref}

\usepackage{bm}
\usepackage{color}
\usepackage{epsfig}

\renewcommand {\phi}{{\varphi}}

\begin{document}
\title{Polarized  edge state emission from  topological spin phases\\ of trapped Rydberg excitons in Cu$_2$O}
\author{A.N. Poddubny and M.M. Glazov}
\affiliation{Ioffe Institute, St. Petersburg 194021, Russia}
\email{poddubny@coherent.ioffe.ru, glazov@coherent.ioffe.ru}
%\date{\today}
\begin{abstract}
In one dimensional chains of trapped Rydberg excitons in cuprous oxide semiconductor the topological spin phase has been recently predicted [Phys. Rev. Lett. {\bf 123}, 126801 (2019)]. This phase is characterized by the diluted antiferromagnetic order of $p$-shell exciton angular momenta{-$1$} and the edge states behaving akin spin-$1/2$ fermions. Here we study the properties of the ground state in the finite chains and its fine structure resulting from the effective interaction of the edge spins. We demonstrate that these edge states can detected optically via the enhancement of the circular polarization of the edge emission as compared with the emission from the bulk. We calculate the distribution of the exciton angular momentum vs. trap number in the chain numerically and analytically based on the variational ansatz.
\end{abstract}
\maketitle 

\section{Introduction}

Search for novel ``quantum'' phases of condensed matter with non-trivial optical, magnetic, and topological properties is among the most vital problems in the field.\cite{Tang:2019aa,Quantum:2020aa} In this regard, interacting quasiparticles in semiconductors and semiconductor nanosystems represent attractive systems. This is because a manipulation of their properties by external fields and tailoring of the environment by nanotechnology offers unique possibility to control  the quasiparticles, their arrangement and interactions on demand. 

A natural quasiparticle formed in semiconductors by means of optical excitation is the exciton, the Coulomb-interaction correlated electron-hole pair.\cite{excitons:RS,ivchenko05a,RevModPhys.90.021001} The large-radius Wannier-Mott excitons %have been 
{were} observed in cuprous oxide {in 1950s}\cite{gross:exciton:eng} and actively studied in various semiconductors and semiconductor-based nanosystems since then. Due to exceptional optical quality of natural Cu$_2$O crystals it has been possible to observe a Rydberg series of excitons up to the principal quantum number $n\gtrsim 25$,\cite{Kazimierczuk2014} and study a plethora of effects inherent to these highly-excited quasiparticles, see Refs.~\onlinecite{Semina2018,Assmann:2020aa} for a review.

The symmetry of conduction and valence bands in Cu$_2$O imposes strict selection rules for the interband optical transitions. In particular, in the dipole approximation the $p$-shell states with the angular momentum (for brevity spin, in what follows) $L=1$ are excited in single-photon processes.\cite{gross:exciton:eng,PhysRev.108.1384,Kazimierczuk2014} Together with the strong van-der-Waals interactions of Rydberg excitons,\cite{Walther2018} it makes possible to study correlated phases of spin-$1$ particles. {The semiconductor host system provides also a controllable and versatile environment highly suitable for optical experiments.\cite{PhysRevB.34.2561,Giessen:conf,Steinhauer:2020aa}}

We have predicted  in Ref.~\onlinecite{PhysRevLett.123.126801} that one-dimensional chains of trapped Rydberg excitons in cuprous oxide have a ground state with a topologically nontrivial spin order, the so-called  Haldane phase.\cite{Haldane1983} This phase is characterized by a diluted antiferromagnetic order and a gapped spectrum of elementary excitations. A specific feature of the Haldane phase is the presence of the edge states which behave as spin-$1/2$ fermions despite the fact that the chain is formed of spin-$1$ particles. 

Naturally, a question arises whether these edge states -- being a hallmark of the Haldane phase -- can, at least in principle, be detected in an experiment. Here we address this question and propose a scheme to detect the edge states optically via the polarization of the emitted light. We demonstrate that the spin state of the exciton in the chain $\bm L$ is transferred to the circular polarization of the emitted light. Namely, the radiative  recombination of the exciton results in the emission of $\sigma^+$- or $\sigma^-$- polarized photon depending on exciton $L_z$ spin component. Importantly, we demonstrate that the unpaired edge spin states result in increase of the $|L_z|$ magnitude at the edges of the chain as compared to its bulk. 

The paper is organized as follows. Section~\ref{sec:model} presents the basics of the model of chains of trapped Rydberg {excitons} introduced in Ref.~\onlinecite{PhysRevLett.123.126801} and a simplified theory for the optical selection rules. Further, Sec.~\ref{sec:res} contains the results of numerical calculations for the finite-length chains of Rydberg excitons and analytical results for a semi-infinite chain. The paper is summarized in a brief conclusion in Sec.~\ref{sec:concl} where the outlook is also presented.

\section{Model}\label{sec:model}

%%%%%%%%%%%%%%%%%%%%%%%%%%%%%%
\begin{figure}[b]
\includegraphics[width=0.45\textwidth]{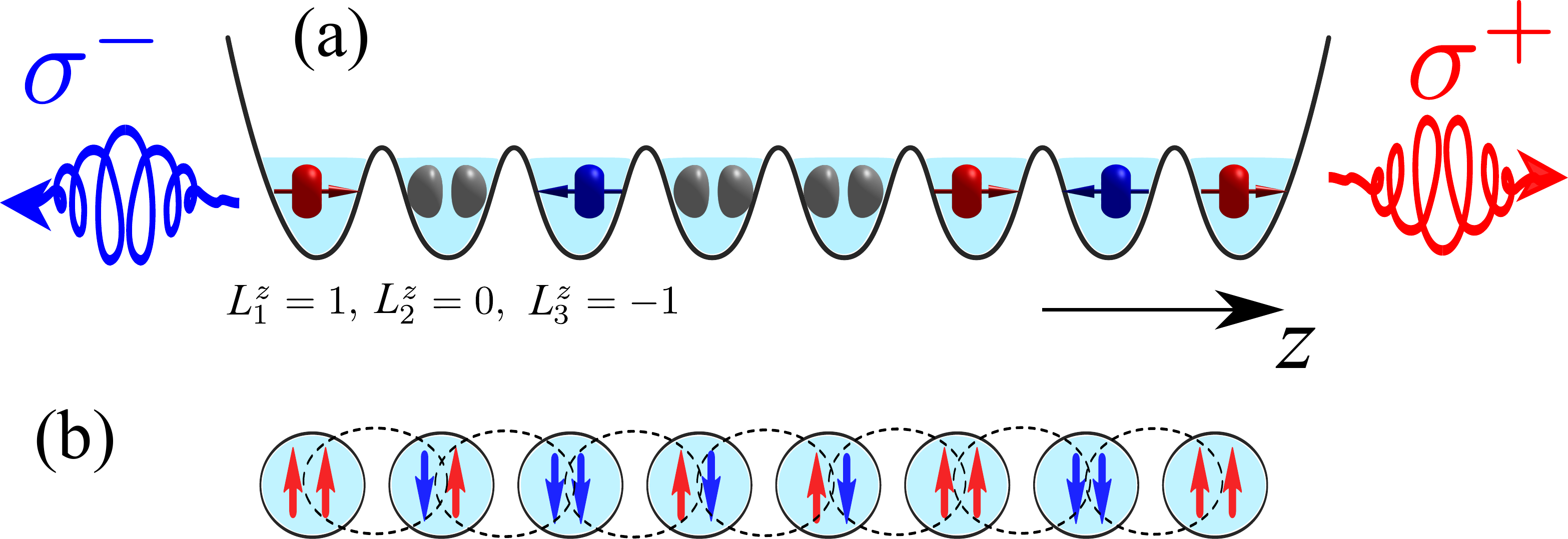}
\caption{ (a) Schematics of the circularly polarized emission from the   diluted antiferromagnetic  state of  $p$-shell Rydberg excitons in an array of traps{.}
(b)  Representation of diluted antiferromagnetic order by a chain of spins-1/2 with singlet coupling{.}
}\label{fig:1}
\end{figure}
%%%%%%%%%%%%%%%%%%%%%%%%%%%%%%

We consider a model system schematically described in Fig.~\ref{fig:1}(a). It consists of one-dimensional periodic array of potential wells which serve as traps for excitons in Cu$_2$O. These traps can be formed by lightwaves akin optical lattices of cold atoms,\cite{Bloch:2005aa}  by strain or electrostatic potentials applied to the semiconductor,\cite{PhysRevB.34.2561} {potential wells formed by focused ion beams\cite{Giessen:conf}} or, alternatively, by trapping excitons in nanocrystals{.}\cite{Steinhauer:2020aa} Let us assume that the traps are sufficiently large to disregard their effect on the electron-hole relative motion envelope $\Phi_{n,l,m}(\bm \rho)$ with $\bm \rho=\bm r_e - \bm r_h$, $\bm r_e$ ($\bm r_h$) being the electron (hole) coordinate, where $n=1,2,3,\ldots$ is the principal quantum number, $l=0,1,2,\ldots, n-1$ is the orbital angular momentum, and $m=-l, \ldots, l$ is its component (or magnetic quantum number). For simplicity we consider isotropic traps where at fixed $n$ and $l$ the $(2l+1)$--fold degeneracy over $m$ remains, see Supplement to Ref.~\onlinecite{PhysRevLett.123.126801} for analysis of the anisotropy and numerical estimates.

In cuprous oxide the topmost valence band and the bottom conduction band are two-fold spin degenerate and have the same parity. For illustrative purposes and simplicity we disregard an actual orbital structure of the valence and conduction band Bloch functions and consider the optical transitions in the minimal model where the angular momentum of light is transferred to the angular momentum of the envelope function. Note that the account for the spin-orbit interaction somewhat modifies the excited states, see Ref.~\onlinecite{efield} and references therein for detail. We, however, disregard this effect and apply the following {simplified} selection rules:
\begin{equation}
\label{sel:simpl}
\begin{cases}
\sigma^+ \to ~~ l=1,~m=1,\\
\sigma^- \to ~~ l=1,~m=-1,
\end{cases}
\end{equation}
where we assumed that the light propagates along the $z$-axis, i.e., the axis of the chain, Fig.~\ref{fig:1}.  In what follows we consider excitons with the same principal quantum number $n$ and $l=1$, which are optically active. {The triplet of the degenerate states with $m=\pm 1, 0$ in the $j$th trap  can be described by the pseudovector angular momentum-$1$ (in what follows, spin-$1$) operator $\bm L_j = (L_j^x,L_j^y, L_j^z)$. Thus, the selection rules~\eqref{sel:simpl} dictate that under the $\sigma^+$ excitation the exciton with $L^z=1$ is generated, while under the $\sigma^-$ excitation the exciton with $L^z=-1$ is formed. Similar rules hold for the exciton recombination.} 

{We note that while intensity of the photoluminescence of excited excitons is controlled by an interplay of the non-radiative relaxation and recombination with the radiative processes, the polarization degree is dictated by the selection rules. For bulk Cu$_2$O crystals with multitude of relaxation and non-radiative recombination pathways for excitons, the  photoluminescence is experimentally studied,\cite{gross:PL,Yu:PL,Gastev:PL,Naka:PL} including the emission of excited $n>2$ excitons.\cite{Gastev:PL,Naka:PL} In structures with traps the exciton energy spectrum becomes discrete and energy conservation law can suppress relaxation processes between discrete levels effectively enhancing the lifetimes of excited states. }

We assume that each trap is occupied by one exciton. The excitons in the neighboring traps are coupled by the van der Waals interaction. The analysis of the relevant energy scales demonstrates that the coupling between the states with different $n$ and $l$ can be neglected.\cite{PhysRevLett.123.126801} Thus, the Hamiltonian of the chain takes the form 
\begin{equation}
\label{chain}
{\mathcal H}=\sum_{j=1}^{N-1} {\mathcal H}_{\rm bond}(\bm L_{j},\bm L_{j+1})\:,
\end{equation}
where $N$ is the number of  traps, $\mathcal H_{\rm bond}$ is the nearest-neighbors interaction Hamiltonian. The general form of ${\mathcal H}_{\rm bond}$ can be established from the symmetry arguments as\cite{Klumper1993}
\begin{align}
\label{Hbond} 
{\mathcal H}_{\rm bond}(\bm L_{1},\bm L_{2})&=c_{0}+c_{1}L_{1}^{z}L_{2}^{z}+c_{2}
(L^{x}_{1} L^{x}_{2}+L^{y}_{1} L^{y}_{2})\\\nonumber &+c_{3}(L_{1}^{z} L_{2}^{z})^{2}+c_{4}
(L^{x}_{1} L^{x}_{2}+L^{y}_{1} L^{y}_{2})^{2}\\\nonumber
&+c_{5}[L_{1}^{z} L_{2}^{z} (L_{1}^{x} L_{2}^{x}+L_{1}^{y} L_{2}^y)+{\rm H.c.}]\\\nonumber&+c_{6}
(L_{1}^{x}L_{2}^{y}-L_{1}^{y}L_{2}^{x})^{2}\:.
\end{align}
Here  $c_0, \ldots, c_6$ are  $7$ real constants {and `H.c.' stands for Hermitian conjugate}. These parameters were calculated in Ref.~\onlinecite{Walther2018} with the result for $n={12\ldots 25}$:
\begin{multline}
c_{0}=-5.58\mathcal E, c_{1}=9.53 \mathcal E, c_{2}=-8.97 \mathcal E, \\c_{3}=1.27\mathcal E, c_{4}=6.59\mathcal E, c_{5}=-3.18\mathcal E, c_{6}=5.04\mathcal E\:.\label{eq:C}
\end{multline}
where the common  factor $\mathcal E$ is ${10^{-4} n^{11}~ }{\rm \hbar\: s^{-1}\times \mu m^{6}}/R^{6}$. Since $c_1>0$ the coupling is mostly antiferromagnetic. Interestingly, that in addition to the rotational symmetry around $z$ axis, the Hamiltonian~\eqref{Hbond} is invariant under the change of parameters $c_{2}\to -c_{2},c_{5}\to -c_{5}$ which corresponds to reflection $z\to -z$ {(i.e., $L^x\to -L^x,~L^y\to -L^y,L^z\to L^z$)} for every second spin. 

In what follows we discuss the ground state of the excitonic chain and its excitations. We demonstrate that the topological edge states can indeed be detected by features of polarization of emission of the $j=1$ and $j=N$ traps.

\section{Results and discussion}\label{sec:res}

%%%%%%%%%%%%%%%%%%%%%%%%%%%%%%
\begin{figure}[tb]
\includegraphics[width=0.45\textwidth]{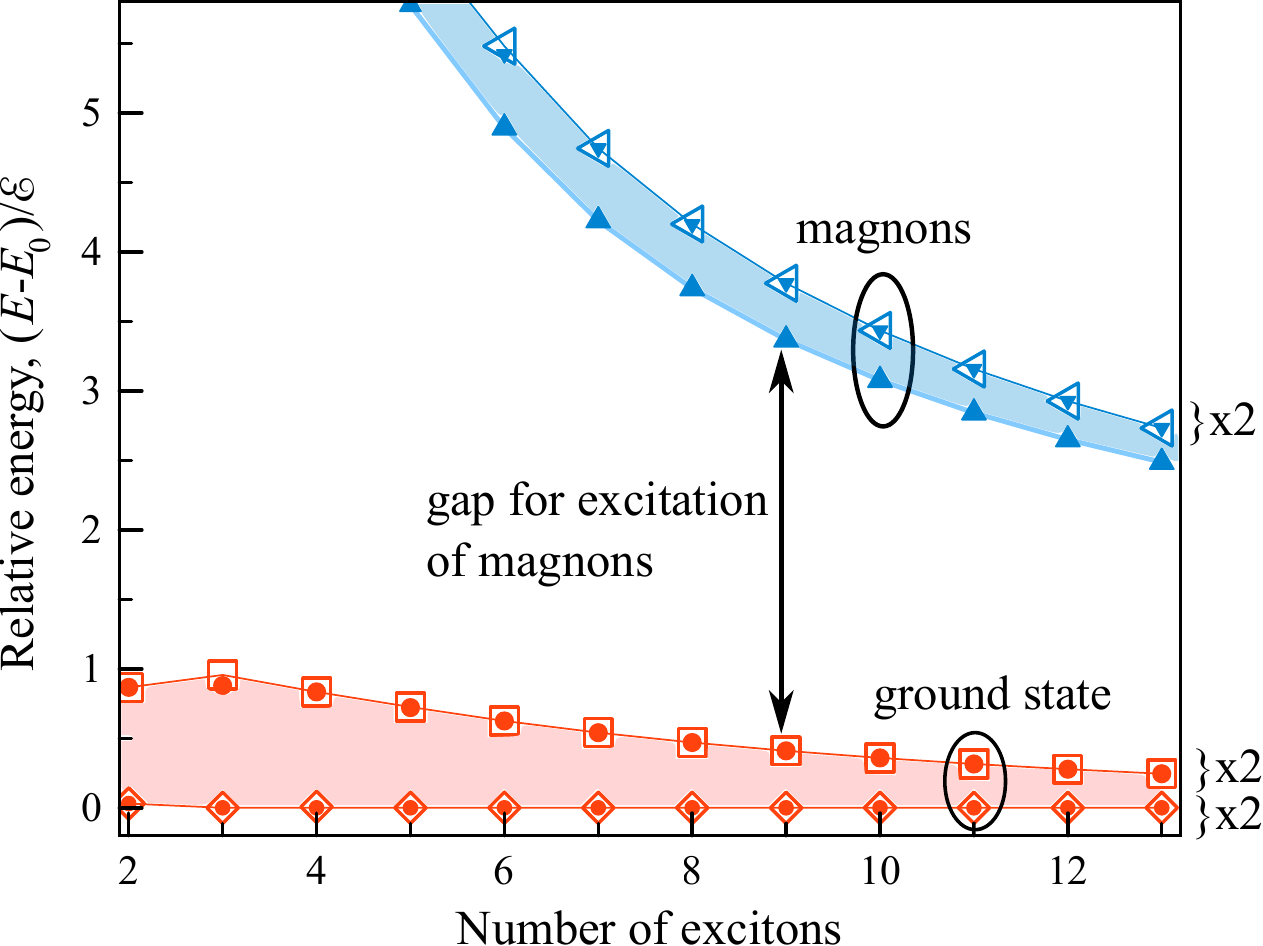}
\caption{ Energy spectrum of the chain with open boundary conditions as function of the number of excitons $N$.
For each $N$ we show only  7 lowest states, energies are counted from the lowest one. Four states converging to the degenerate ground state in the limit of infinite chain are shown by red squares and circles. Three magnon excitations with $S_z=0$ and $S_z=\pm 1$ are shown by blue triangles. 
}\label{fig:2}
\end{figure}
%%%%%%%%%%%%%%%%%%%%%%%%%%%%%%

Figure~\ref{fig:2} shows the energy of seven lowest states in the finite array of traps with open boundary conditions, depending on the number of excitons $N$ {found {by numerical} diagonalization of the Hamiltonian~\eqref{chain}. In our calculations, we have used the energy of the ground state\cite{PhysRevLett.123.126801}
\begin{equation}
\label{ground}
E_0 \approx -3.51N\mathcal E
\end{equation}
as the origin of the energy.} 
Red circles and squares correspond to 4 lowest states. {The analysis performed in  Ref.~\onlinecite{PhysRevLett.123.126801} based on the  infinite time-evolving block decimation (ITEBD)\cite{Vidal2003,Vidal2004,Vidal2007} and Kennedy-Tasaki\cite{Kennedy1992b} variational ansatz demonstrates that, i}n the limit of infinite number of excitons, $N\to \infty$, these states become degenerate and form 4-fold degenerate ground state of the Haldane model.\cite{AKLT,affleck1988} The remaining three states with higher energy, shown by blue triangles,  correspond to the single-magnon  excitations on top of these 4 lowest states.~\cite{PhysRevLett.123.126801,AROVAS1989,Bartel2003} The magnon can be characterized by the projection of the total orbital momentum
on the structure axis and form a singlet with ${S}_z=0$ and a degenerate doublet with ${S}_z=\pm 1$. {In the limit of $N\to \infty$ the magnon excitation gap tends to $\approx 1.1\mathcal E$.\cite{PhysRevLett.123.126801} Substantial dependence of the magnon energy on the chain length observed in Fig.~\ref{fig:2} is related to the size quantization effect: {T}he magnons form standing waves with the wavevector determined by the reciprocal chain length.}The calculation demonstrates the decrease of the energy gap between {the} 4 lowest states and the magnons with the increase of the number of excitons because  the size quantization energy of magnons  becomes smaller.

We will now examine in more detail the spin structure of 4 ``ground'' states of the exciton {chain} and show, how it reflects the diluted antiferromagnetic order in the system.  Our goal is to demonstrate that (i) these 4 states behave as  spin-1/2 topological edge states of the Haldane phase and {(ii) they} can be probed {optically} by studying the circular polarization of the exciton emission from the traps given by $\langle L_j^z\rangle$. 

First, we note that two interacting spins 1/2 at the edges form a degenerate doublet with the total momentum projection on $z$ axis $L_z=\pm 1$ 
and two almost degenerate states with $L_z=0$. {Indeed the axial symmetry of the system together with the time-reversal invariance ensures the degeneracy of $L_z$ and $-L_z$ states, but enables mixing the states with the same angular momentum component. The latter coupling vanishes in $N\to \infty$ limit because it is related to the finite localization length of the edge states {[cf. Ref.~\onlinecite{PhysRevLett.125.056401} where the coupling between the edge spins in the nanotube is studied].} This explains  the separation of  the 4 lowest states in Fig.~\ref{fig:2} into 2 doublets. The calculation shows that the pair with $L_z=0$ has lower energy for even values of  $N$ and the pair with $L_z=1$ has lower energy for odd $N$. 

%%%%%%%%%%%%%%%%%%%%%%%%%%%%%%
\begin{figure}[ht]
\includegraphics[width=0.45\textwidth]{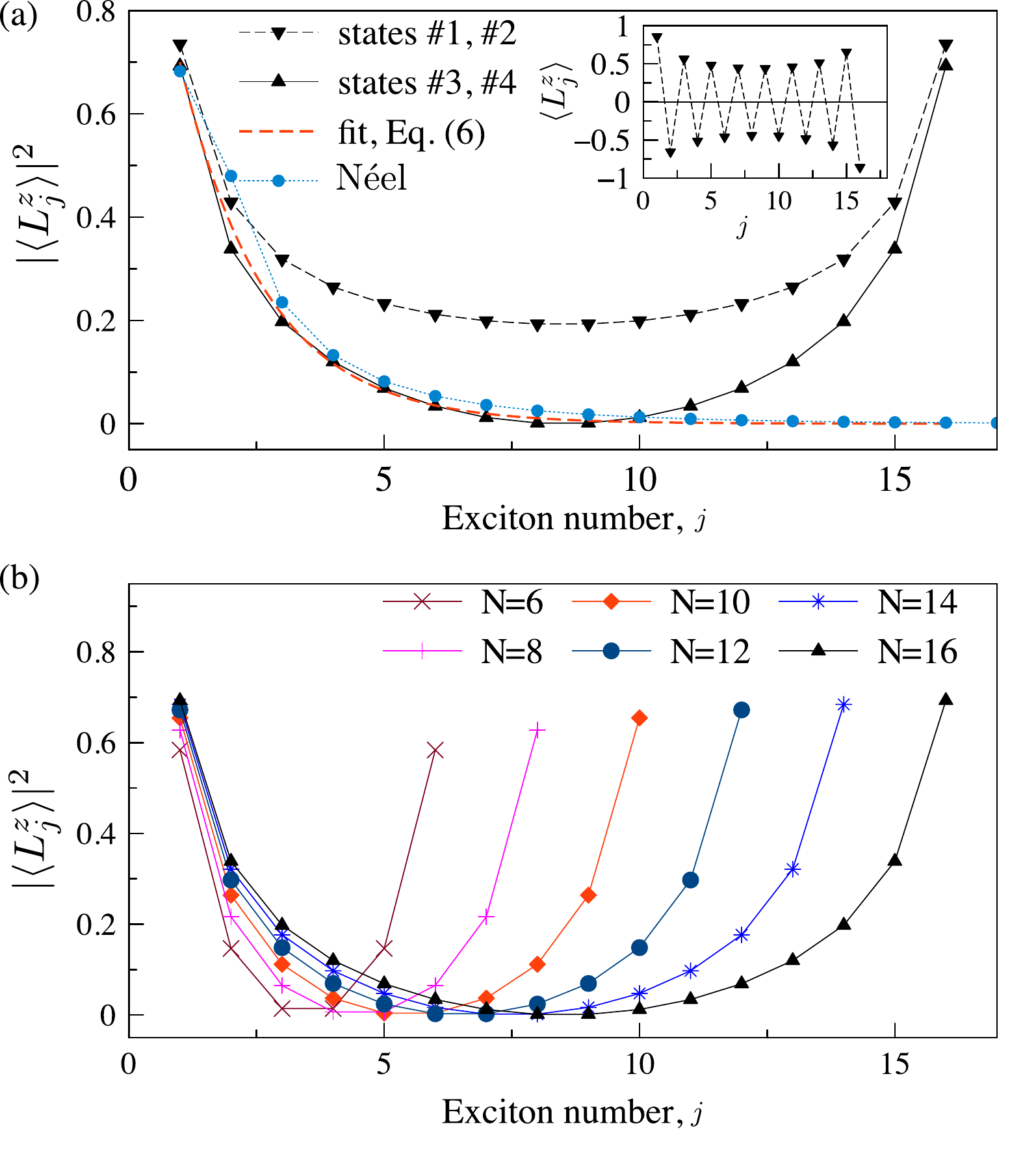}
\caption{(a) Spatial distribution of the angular momentum $\langle L_j^z \rangle$ for two lowest states (downward-pointing triangle) and two second two lowest states (upward-pointing triangle). Inset shows the dependence $\langle L_j^z \rangle$ for the lowest state.
Red curve shows the fit Eq.~\eqref{eq:an}. {Blue circles show the decay of the N\'eel correlator squared $|C_{\text{N\'eel}}(j)|^2$, Eq.~\eqref{neel}, calculated for the infinite chain via ITEBD approach, see text for details.} {Chain length is $N=16$.}
(b) Spatial distribution of the angular momentum $\langle L_j^z\rangle ^2$ for the third state depending on the number of excitons in a chain.
}\label{fig:3}
\end{figure}
%%%%%%%%%%%%%%%%%%%%%%%%%%%%%%

Second,  we have {presented} in  Fig.~\ref{fig:3}
the spatial distribution of the average value $\langle L^z_j\rangle$ along the {chain} calculated {numerically} for the lowest 4 states of the system. Figure~\ref{fig:3}(a) demonstrates that for both states the spin $\langle L^z_j\rangle^2$  is at maximum {at} the edges and decays to the center. This is in full agreement with that could be  expected from the edge states. The sign of $\langle L^z_j\rangle$ oscillates  due to the antiferromagnetic order, as demonstrated by the inset of   Fig.~\ref{fig:3}(a). 
Figure~\ref{fig:3}(b) shows the dependence of the spin distribution on the number of {traps in the chain}. The calculation demonstrates that the distribution has apparent edge maxima already for $N=6$ excitons (brown crosses). When the number of excitons increases {up to} 10, the spins are mostly concentrated at the edges, where $\langle L^z_j\rangle^2\approx 0.7$ and the values of $\langle L^z_j\rangle^2$ in the middle of the array are smaller by more than an order of magnitude. {For sufficiently large $N$} %In the limit of semi-infinite structure 
the distribution of spin can be approximately described by the {exponentially} decaying function 
\begin{equation}\label{eq:an}
\langle L^z_j\rangle {=} L^z_0 a^{j},\quad j=1,2\ldots. 
\end{equation}
The corresponding fit for {the} 3rd and 4th states {at $N=16$} with  $L_0 \approx 1.36 $, $a\approx -0.75$ is shown by the red dashed curve in Fig.~\ref{fig:3}(a). The decay in the figure is twice faster because  the $\langle L^z_j\rangle^2$ dependence on the site number $j$ is shown. There exists also a noticeable difference between the decays of polarization for the pairs of states 1,2 and 3,4. Most probably, it is a finite size effect beyond the scope of our analysis.

{The exponential decay in Eq.~\eqref{eq:an} calls for special analysis. We demonstrate now that  Eq.}~\eqref{eq:an} follows from  the variational matrix product state ansatz for the wavefunction,\cite{Orus2014}
 \begin{equation}\label{eq:mps1}
\psi_{\alpha\beta}^{s_1\ldots s_N}=\mathbb M_{\alpha\alpha_1}^{s_1}\mathbb M_{\alpha_1\alpha_2}^{s_2}\ldots \mathbb M_{\alpha_{N-1}\beta}^{s_N},
\end{equation}
where the rank-$2$ matrices $\mathbb M^{s_j}$ ($s_j=-1,0,1$) are given by
\begin{gather}
\label{eq:matrices}
\mathbb M^1=\frac1{\sqrt{x^2+1}}\begin{pmatrix}
0&0\\x&0
\end{pmatrix},\\\nonumber \mathbb M^{-1}=\frac1{\sqrt{x^2+1}}\begin{pmatrix}
0&-x\\0&0
\end{pmatrix},\mathbb M^{0}=\frac1{\sqrt{x^2+1}}\begin{pmatrix}
-1&0\\0&1
\end{pmatrix}.
\end{gather}
Here the subscripts $\alpha,\beta=1,2$ depend on the specific boundary condition and the superscripts $s_j=-1,0,1$ label the spin projections.
The matrices~\eqref{eq:matrices} and, accordingly, the wavefunction~\eqref{eq:mps1} depend on a single variational parameter  $x$ that determines the average value of spin in the infinite array with the periodic boundary conditions, 
\begin{equation}\label{eq:Lz2}
\langle L_z^2\rangle=\frac{x^2}{x^2+1}\:.
\end{equation}
Physically, this parameter characterizes the anisotropy of the spin distribution. {The cases of $x=0,\infty$ correspond to the extreme anisotropy:} Obviously, for $x=0$ one has  $L_z=0$ for all the spins. {For $x\to \infty$ one has $|L_z|=1$. The latter also corresponds to a} strongly anisotropic model where $L_z=\pm 1$ and the value $L_z=0$ is impossible, i.e., is much higher in energy.
In case of isotropic Affleck-Kennedy-Lieb-Tasaki (AKLT) model, \cite{AKLT,affleck1988,Kennedy1992b} where the Hamiltonian
${\mathcal H_{\rm bond}}= {\bm L}_{1}\cdot{\bm L}_{2}+({\bm L}_{1}\cdot{\bm L}_{2})^{2}/3$, \cite{}
the ansatz Eq.~\eqref{eq:mps1} is exact with  $x=\sqrt{2}$.  

For our case of Rydberg excitons in Cu$_2$O the ground state energy in the limit of $N\to \infty$ {calculated variationally with the trial function~\eqref{eq:mps1} parametrized via the matrices $\mathbb M^{s}$ in Eq.~\eqref{eq:matrices}} has a minimum with  $E_0 \approx -3.49N\mathcal E$  at    $x\approx 2$.\cite{PhysRevLett.123.126801} This variational result is in excellent agreement with the numerical result Eq.~\eqref{ground}. The expectation value of the operator $L^z_j$ in a chain with open boundary conditions can be then found from the ansatz \eqref{eq:mps1} as
\begin{equation}\label{eq:expectation1}
\langle\psi_{\alpha\beta}|L^z_j|\psi_{\alpha\beta}\rangle=
[\underbrace{\mathbb I\cdot \mathbb I\cdot \ldots}_{j-1}\cdot\mathbb L_z\cdot \underbrace{\ldots\cdot\mathbb I\cdot\mathbb I}_{N-j}]_{\alpha\beta},
\end{equation}
where 
\begin{equation}\label{eq:I}
\mathbb I=\frac{1}{x^2+1}\begin{pmatrix}
1&x^2\\x^2&1
\end{pmatrix},\quad 
\mathbb L_z=\frac{1}{x^2+1}\begin{pmatrix}
0&-x^2\\x^2&0
\end{pmatrix}\:.
\end{equation}
Taking into account that the matrix $\mathbb I$ has eigenvalues $1$ and $(1-x^2)/(1+x^2)$, we recover the decay law  Eq.~\eqref{eq:an}. {Note, that the expression~\eqref{eq:expectation1} does not depend on $\alpha,\beta$ for $j>2$.}
The ansatz Eq.~\eqref{eq:mps1} predicts a somewhat faster decay with 
\begin{equation}\label{eq:a}
a=\frac{1-x^2}{1+x^2}\approx -0.6
\end{equation}
than the value $a\approx -0.75$ following from the fit of exact numerical result in Fig.~\ref{fig:3}(a). Better  accuracy can be obtained if the number of variational parameters is increased, i.e. the rank of the matrix $\mathbb M^s$ is chosen larger than 2.

An even more intuitive way to understand  the origin of edge spin states is provided by the construction, where each spin-$1$ particle  is {considered as a composite particle made of} two spin-$1/2$   fermions. The spin-$1$ state is realized by  the three triplet states, as shown in Fig.~\ref{fig:1}(b), \cite{AKLT} while the singlet state with the total spin $0$ is assumed to be split by a significant energy.
Namely,  we represent the $L_z=1$ state by two spins $\uparrow\uparrow$, the state $L_z=-1$ by two spins $ \downarrow\downarrow$, and 
$L_z=0$ by a symmetric combination of $\downarrow\uparrow$ and  $\uparrow\downarrow$.  Next, in agreement with the general approach for the description of Haldane phase\cite{AKLT,affleck1988,Nijs1989} we assume
 that spins 1/2 on adjacent sites should always be in the singlet states, i.e. oppositely oriented, e.g.
$( \uparrow\uparrow)(\downarrow\downarrow)$ or $( \uparrow\uparrow)(\downarrow\uparrow)${, Fig.~\ref{fig:1}(b)}. Such fusion rules {for spins at the neigbouring sites} automatically ensure the long-range diluted antiferromagnetic order {realized in our exciton spin chains:\cite{PhysRevLett.123.126801} T}he state with $L_z=1$ is always followed by $L_z=-1$ after {an arbitrary %some possible 
number $n_0=0,1,2,\ldots$ of states with} $L_z=0$.  
{One can readily see that t}his construction leaves two uncoupled  spins-1/2 at the structure edges, that are responsible for formation of edge states. This construction also  allows us to re-derive the spin decay law Eq.~\eqref{eq:an}. Namely, we set the probabilities of each of the configurations $\uparrow\uparrow$ and $\downarrow\downarrow$ to $p_1/2$ with
\[
p_1=\frac{x^2}{x^2+1}\:,
\]
and the probabilities of each of the configurations $\uparrow\downarrow$ and $\downarrow\uparrow$ to $p_2/2$ with
\[
p_2=\frac{1}{x^2+1},
\]
in order to satisfy Eq.~\eqref{eq:Lz2} for the  average spin. The normalization condition reads ${p_1+p_2}=1$.  These probabilities make it possible to determine the law of the $\langle L_j^z\rangle$ decay. To that end,
 without the loss of generality we consider the situation when the first spin-1/2 at the first site is fixed to $\uparrow$, i.e. the chain starts either  with
$(\uparrow\uparrow)$ (with the probability $p_1$) or with $\uparrow\downarrow$  (with the probability $p_2=1-p_1$). Hence, the average spin projection at the first site is equal to
$
\langle L^z_1\rangle=p_1\times 1+p_2\times 0=p_1.
$
The average spin at the second site is then contributed only by configurations with $L_z^{(2)}\ne 0$, namely
$
 (\uparrow\uparrow)(\downarrow\downarrow)$ with the probability $p_1^2$ and $L_z^{(2)}=-1$
and
$(\uparrow\downarrow)(\uparrow\uparrow)$  with the probability $p_2p_1$ and $L_z^{(2)}=+1$\:.
%\]
Hence, 
\[
{\frac{\langle L^z_2\rangle}{\langle L^z_1\rangle}
=\frac{1}{p_1} \left[p_1\times (-1)+p_2p_1{\times (+1)}\right]}=\frac{1-x^2}{1+x^2},
\]
in agreement with Eq.~\eqref{eq:a}. {Furthermore, repeating this procedure we obtain the same ratio for $\langle L^z_{j+1}\rangle/\langle L^z_{j}\rangle\equiv a$. 
}

{Interestingly, the decay of the edge spin polarization $L_j^z$ well matches the behavior of the N\'eel correlator
\begin{equation}
\label{neel}
C_{\text{N\'eel}}(j) = (-1)^j\langle L_{i}^z L_{i+j}^z\rangle,
\end{equation} 
in the infinite chain. Here, the averaging over the trap $i$ at a fixed $j$ is assumed. The quantity $|C_{\text{N\'eel}}(j)|^2$ calculated via the ITEBD approach\cite{PhysRevLett.123.126801} is shown by the blue circles in Fig.~\ref{fig:3}(a), being in good agreement with the results of calculated decay of the ${\langle}L_j^z{\rangle}^2$ in $N=16$ chain. This is not surprising because the decay of topological edge states at $j\gg 1$ is determined by the bulk properties, namely, by the spin-spin correlation function. It is particularly clear in the variational ansatz{,} Eq.~\eqref{eq:mps1}, which shows that (up to the inessential details related to the choice of the initial spin state and the common factor) the N\'eel correlator Eq.~\eqref{neel} and the $L_j^z$ are given by the same construction, Eq.~\eqref{eq:expectation1}.}

{As a result, the topological edge states provide enhanced values of $L^z_1$ and $L^z_N$ in the chains of trapped Rydberg excitons. This effect can be  detected optically via the enhanced circular polarization of the emission in accordance with selection rules discussed in Sec.~\ref{sec:model}, Eq.~\eqref{sel:simpl}.}

{It is worth to note, that high values of edge spins $L^z_1$ and $L^z_N$ in the finite-size chains as compared to the values in the bulk present a fingerprint of the diluted antiferromagnetic -- Haldane -- phase studied here. We have checked numerically that for the chains with the interaction parameters corresponding to the real ferromagnetic or antiferromagnetic order (see Supplement in Ref.~\onlinecite{PhysRevLett.123.126801})  the absolute value $|{\langle} L_j^z{\rangle}|=const$. In the latter situation the fusion rules discussed above are not applicable.}

\section{Conclusion and outlook}\label{sec:concl}

Here we have studied the distribution of $p$-exciton angular momentum in the one-dimensional chains of trapped excitons in Cu$_2$O. We have shown that the predicted topological Haldane phase with diluted antiferromagnetic order and four-fold degenerate ground state behaving as two spin-$1/2$ fermions at the edges manifests itself in the circular polarization of the exciton emission. The four-fold degeneracy of the ground state is lifted in the finite chains. We have demonstrated that the exciton angular momentum component ${\langle}L^z{\rangle}$ at the edges is enhanced as compared to the bulk value. We have calculated the distribution of the angular momentum vs. trap position in the chain numerically and also analytically within two approaches: (i) variational one based on matrix-product-state ansatz and (ii) two spin-$1/2$ representation of the angular momentum-$1$ state with appropriate ordering rules. In particular, these auxiliary -- fictitious -- spins-$1/2$ in the adjacent traps are always antiparallel, while at the same trap they form one of the triplet states. {It is of interest to address the processes of thermalization of the spin degrees of freedom in excitonic chains and analyze  an interplay of the recombination and relaxation processes. This is a problem for a separate study.}

In cuprous oxide studied here the formation of the topological spin (strictly speaking, orbital angular momentum) states is caused by particular relation between the van der Waals interaction parameters. It could be interesting to analyze the situation where the excitons have orbital angular momentum $0$, but the exchange interaction between the electron and hole provides spin-$1$ state (like, e.g., in GaAs-, CdTe-, or transition-metal dichalcogenides-based nanosystems). In this situation, one can imagine similar effects if the spin-spin coupling between the adjacent traps is sufficiently strong. Here, however, the spins-$1/2$ are real as they correspond to the electron and hole spins (or valley indices) forming the exciton. Studies of such systems can shed more light on the topological properties of spin-$1$ chains.

%%%%%%%%%%%%%%
%\emph{Acknowledgements.} 
\acknowledgements
We are grateful to M. A\ss mann, M. Bayer, T. Pohl, M.A. Semina, and V. Walther for fruitful discussions. MMG is grateful to RSF Project No. 17-12-01265 for partial support.
ANP acknowledges support of the Russian President Grant No.~MD-243.2020.2.

%\nocite{apsrev41Control}
%\bibliographystyle{apsrev4}
%
%\bibliography{titleon,Cu2O}

\end{document}